\documentclass{article}

\usepackage[pdftex]{graphicx}

\begin{document} 

\title{Optical Approach to Gravitational Redshift} 

\author{Yong Gwan Yi} 

\maketitle

\begin{abstract} 

An optical approach begins by interpreting the gravitational 
redshift resulting to a change in the relative velocity of light due to the medium of propagation in the gravitational field. 
The discussion continues by pointing out an agreement in structure between the equation for rays in geometrical optics and the geodesic equation of general relativity. From their comparison we learn that the path of rays should be given by $ds^2=n^2(r)dr^2+r^2d\theta^2$, not by $ds^2=dr^2+r^2d\theta^2$, in a medium with spherical symmetry of refractive index $n(r)$. The development of an optical analogy suggests introducing $n^2(r)$ in place of $g_{rr}$ as an optical version of the Schwarzschild metric. In form and content, $n^2(r)$ is different from $g_{rr}$. The optical point of view replaces the general-relativity explanations in terms of time and gravitation.

\end{abstract}

\bigskip 

\section{Introduction} 

Four classic tests are recognized as experimental verifications of the general theory of relativity: the gravitational redshift of spectral lines, the deflection of light by the Sun, the precession of the perihelion of the orbit of the planet Mercury, and the time delay of radar echoes passing close to the Sun. Three of these tests examine the influence of the gravitational potential on the propagation of light. Only the planetary orbit 
precession investigates the motion of a particle of finite mass in the gravitational field of the Sun. The general theory has been essentially mathematical in character, being concerned with the consequences of a ``geometrization'' of the space-time manifold. The application of the theory involves thus the use 
of special mathematical methods which, although relevant to optics in three cases, may easily be considered separately from it. Because these are optical phenomena, nevertheless, one may raise a question as to whether the three classic tests can also be inferred from the point of view of optics.

Before the general theory of relativity, Einstein (1911) himself inferred the redshift effect from a consideration of the velocity of light in the gravitational field, and derived the deflection of light in the solar gravitational field by means of Huyghens' principle. Eddington (1920), who verified experimentally the bending of light rays round the Sun during the eclipse in 1919, calculated the bending effect by assuming a gravitational index of refraction $n(r)=1/(1-2GM_{\odot}/c^2r)$. The idea of an equivalent refractive medium was utilized in the investigation of specific problems, but the effective index of refraction was defined by the isotropic form of the Schwarzschild metric (Plebanski 1960; de Felice 1971). For educational purpose, Schiff (1960) showed how the deflection of light can correctly be inferred from Huyghens' principle. The optical analogy gives the results in a way that is easy and instructive. The problem is that the discussion and results do not go beyond an formal analogy of the classic tests. Their formal analogy cannot be an answer to the question. 
   
An optical approach begins by interpreting the gravitational 
redshift resulting to a change in the relative velocity of light due to the medium of propagation in the gravitational field. From the optical point of view it is natural to think of the effect as due to the velocity change of light in the medium of propagation. In such a thought, the redshift effect has nothing to do with relativity but rather is of a purely optical phenomenon. When we check an optical form of expression for the deflection of light, we find that the equation of rays has previously had the form as given by the geodesic equation. It is of particular interest to note the fact that there is an agreement in form between the equation for rays in geometrical optics and the geodesic equation of general relativity. It becomes evident that the classic tests can also be predicted in exactly the same form from the optical point of view. The development of an optical analogy suggests introducing an optical version of the Schwarzschild metric. In this attempt to deduce their optical nature, one may pass from the mathematical language to the physical language, and see how they can be reconciled with each other.

\section{Gravitational Redshift} 

In 1911, Einstein predicted a change in the frequency of spectral 
lines with gravitational potential, generally referred to as the 
gravitational redshift. His argument begins by pointing out the equivalence between a system in the gravitational field of ${\bf g}$ and a gravitation-free system with acceleration $-{\bf g}$. From the customary point of view, however, it is natural to think of the frequency change as a result of the relative velocity change of light. As can also be seen in his argument, phenomenologically at least, the effect would appear to be due to the change of velocity which the radiation experiences during propagation along the lines of force of the gravitational field. If we try to find an optical interpretation of this change of velocity, we find without difficulty that it is equal to the velocity difference due to the medium which the radiation experiences during propagation. 

Let $\rho$ be the density and $\phi$ be the gravitational 
potential with ${\bf g}=-\nabla\phi$. A pressure uniform 
throughout a fluid mass produces no effect on the motion. 
The time rate of change of the momentum of fluid is equal to 
and opposite to the pressure gradient force in the medium. 
The velocity difference due to fluid with differing 
gravitational potential is calculated according to 
\begin{equation} 
\frac{d{\bf v}}{dt}=-{\bf g}\quad\mbox{or}\quad 
\nabla\phi. 
\end{equation} 
This leads to a phenomenological interpretation of it: the 
redshift effect is attributed to the relative velocity change 
due to the medium by which light is affected during propagation. It is natural to have an intuitive way of arriving at the redshift result. In such an intuitive way, the redshift effect appears to be an optical phenomenon in nature. There is no difference in the redshift result from Einstein's argument. There is thus no objection to interpreting the redshift effect from the optical point of view.

A difference of interpretation already existed at the first time 
of observation. Jewell in 1897 and Fabry and Boisson in 1909 
found displacements of solar spectral lines toward the red 
end of the spectrum, and ascribed them to an effect of 
pressure in the absorbing layer. However, Einstein's 
theory of general relativity in 1916 established in most 
physicist's minds the interpretation of redshift as a 
manifestation of time dilation in a gravitational potential. 
This interpretation has been regarded as justified in the context of the space-time formulation of general relativity. But when viewed from the optical point, phenomenologically, the redshift is no more than an effect that the velocity of light is altered, linearly to a first approximation, by the medium as a result of pressure gradient force. The optical point of view reopens the question of interpretation and reminds us of the effect of pressure on the redshift of solar spectral lines. This means that, not only the gravitational potential, but any change in mechanical pressure, density and temperature of the medium would give rise to an effect of the same kind on the redshift of spectral lines. 

In order to complete the present description, it is 
necessary to consider the hydrodynamic equation. 
The hydrodynamic equation is 
\begin{equation} 
\rho\frac{d{\bf v}}{dt}=-\rho{\bf g}-\nabla P+{\bf F}_{vis}, 
\end{equation} 
where $\nabla P$ is the gradient of pressure and ${\bf F}_{vis}$ is viscous force. In case of the conducting medium with 
a magnetic field, it is necessary to include the magnetic force 
term in the hydrodynamic equation. We restrict our discussion 
to one-dimensional case along the lines of force of gravitational 
field. If we neglect viscous effects, the hydrodynamic equation 
takes the form 
\begin{equation} 
\rho\frac{dv}{dt}=\rho\frac{d\phi}{dr}-\frac{dP}{dr}. 
\end{equation} 
Since the time required to propagate a path $dr$ is $dr/c$ 
to a first approximation, the integration of (3) gives 
\begin{equation} 
\triangle v=\frac 1c\biggl(\triangle\phi- 
\frac{\triangle P}{\rho}\biggr) 
\end{equation} 
for the relative change of velocity in the medium which light 
experiences during propagation along the path. The velocity 
of light at the point of observation thereby becomes 
\begin{equation}
c'(r)=c\biggl[1+\frac{\triangle v}{c}\biggr]= 
c\biggl[1+\frac{1}{c^2}\biggl(\triangle\phi- 
\frac{\triangle P}{\rho}\biggr)\biggr], 
\end{equation} 
as compared with its velocity $c$ at the moment of emission. 
By Doppler's principle, it can be written in terms of frequency as 
\begin{equation} 
f'(r)=f\biggl[1+\frac{1}{c^2} 
\biggl(\triangle\phi-\frac{\triangle P}{\rho}\biggr)\biggr]. 
\end{equation} 

The optical approach sheds additional light on its relation to property of the medium of propagation. The speed of light in (5) can be thought of as a speed in a medium with an index of refraction given by $c'(r)=c/n(r)$. Thus 
\begin{equation} 
\frac{1}{n(r)}=1+\frac{1}{c^2}\biggl(\triangle\phi- 
\frac{\triangle P}{\rho}\biggr). 
\end{equation} 
This illustrates how the gravitational redshift can be understood in terms of a refractive index of the medium of propagation. It has a consequence which is of fundamental importance for describing the deflection of light and the radar echo delay from the point of view of optics. To reconcile optics and general relativity, however, a vacuum must be understood to exclude even a gravitational potential.

The redshift effect was qualitatively in agreement with 
astronomical observations both in the case of the Sun and in 
the case of white dwarf star like Sirius B where the effect is 
about thirty times larger. However, the quantitative agreement 
was not very good. While the predicted shift is independent 
of the point of observation on the solar disk, observations have shown that the wavelength of spectral lines increases as the point of observation moves toward the limb (Higgs 1960). Furthermore, the solar lines observed at the limb are definitely asymmetric, having pronounced red flanks. There seems to be a systematic change in profile as one approaches the limb. In atomic spectra (White 1934), the broadening of a spectrum line due to pressure has shown that the spectrum line observed is spread out more on the long wavelength side than it is on the short. With increasing pressure, the mean collision time increases and the time between collisions decreases with the result that, as the line is shifted to the red, it is broadened asymmetrically. From this point of view, the asymmetry observed in limb lines seems to be of pressure character. In fact, Blamont and Roddier (1961) found a complete interpretation of their experimental value at the limb when they added to the gravitational redshift the pressure redshift of the Lindholm effect. Together with asymmetric profile, their interpretation reminds us of the effect of pressure on the redshift of solar spectral lines. 

The controlled experiments using M\"{o}ssbauer effect 
are able to test the gravitational redshift to an excellent 
accuracy. In the experiments, $\gamma-$rays in a nuclear 
resonance passed through an evacuated tube or a tube filled 
with helium along the lines of force of the gravitational field. 
The redshifts observed by Pound and Rebka (1959, 1960) were shown to be in agreement with the predicted shift when the effects were corrected by the temperature difference between source and observer. A measurement of redshifts in a rapidly rotating system was shown to fit the dependence of an effective acceleration of gravity on angular velocity (Hay et al. 1960). 

\section{Deflection of Light} 

The Schwarzschild metric, appropriate for the region 
exterior to a spherically symmetric distribution of 
mass $M$, is given in the standard form as 
\begin{equation} 
c^2d\tau^2=\biggl(1-\frac{2GM}{c^2r}\biggr)c^2dt^2- 
\biggl(1-\frac{2GM}{c^2r}\biggr)^{-1}dr^2\\-r^2 
(d\theta^2+\sin^2\theta d\varphi^2). 
\end{equation} 
In what follows, we use for the components of the 
metric tensor the expressions 
$g_{00}(r)=1/g_{rr}(r)=1-2GM/c^2r$. Assuming 
that the whole motion takes place in the plane 
$\varphi=0$, we obtain as the equations of motion 
three differential equations. For light rays 
propagating along the geodesic lines, we replace 
the parameter $\tau$ by a parameter $p$ describing 
trajectory. In particular, one may choose to 
normalize $p$ so that 
\begin{equation} 
g_{00}\frac{cdt}{dp}=1\quad\mbox{in}\quad 
g_{00}\frac{cdt}{dp}=\mbox{constant}. 
\end{equation} 
On this normalization condition the three 
differential equations can be combined into 
one differential equation, which is of the 
same structure as (8). According to Weinberg (1972), 
the change in $\theta$ as $r$ decreases from 
infinity to its minimum value 
$r_0$ is given by 
\begin{equation} 
\triangle\theta=\int\limits_{\infty}^{r_0} 
\biggl[\frac{g_{00}(r_0)r^2}{g_{00}(r)r_0^2}-1 
\biggr]^{-1/2}\frac{g_{rr}^{1/2}(r)dr}{r}. 
\end{equation} 
This integral can be evaluated by expanding in the 
small parameters $GM/c^2r$ and $GM/c^2r_0$ to first 
order, giving $4GM/c^2r_0$ for a light ray deflected 
by the Sun. 

Let us consider rays in a medium which has spherical symmetry, 
i.e. where the refractive index depends only on the distance 
$r$ from a fixed point O: $n=n(r)$. This case is approximately 
realized by the Earth's atmosphere, when the curvature of the 
Earth is taken into account. The light rays are then plane curves, situated in a plane through the origin. If $(r,\theta)$ are the polar coordinates of a plane curve, the angle $\psi$ between the radius vector to a point $r$ on the curve and the tangent at $r$ is given by 
\begin{equation} 
\sin\psi=\frac{rd\theta}{(dr^2+r^2d\theta^2)^{1/2}}. 
\end{equation}
Along each ray  
\begin{equation} 
n(r)r\sin\psi=\mbox{constant}. 
\end{equation}
The relation in (12) represents a law of refraction in a medium which has spherical symmetry. Actually, Snell's law of refraction becomes the form of (12) when applied to a medium of spherically varying refractive index.\footnote{Snell's law can be written $n(r)\sin\psi=n(r+dr)\sin(\psi+d\psi)$, from which we obtain $d\psi=-(n/dn)\tan\psi$, where $\tan\psi=rd\theta/dr$. In order to give an angle of refraction from the spherically varying radius vector, we must add $d\theta$ to the angle $d\psi$ between the incident at $r$ and refracted ray at $r+dr$. So we have $d\psi+d\theta=-(n/dn)\tan\psi$, obtaining the relation in (12).}   
Since $\psi=\pi/2$ at the point $r_0$ of closest approach of the ray to the origin, (12) may also be written as $n(r_0)r_0=$ constant. This relation is sometimes called the formula of Bouguer in geometrical optics. The law of refraction can therefore be written in the form
\begin{equation} 
\triangle\theta=\int\biggl[\frac{n^2(r)r^2}{n^2(r_0)r_0^2}-1 
\biggr]^{-1/2}\frac{dr}{r}, 
\end{equation} 
which is the equation of rays in a medium with spherical symmetry (Born and Wolf 1975).

At first sight, we can see a striking equivalence of expressions 
between the equation for rays in geometrical optics and the 
geodesic equation expressed in (10). Compared to the geodesic 
equation, the equation for rays is lacking a term arising 
from the difference in path length. For lack of the term, the equation of rays corresponds to the case which is obtained 
when the curvature of the physical space is neglected. According to general relativity, the deflection of light is due partly to 
the varying velocity of light and partly to the non-Euclidean 
character of the spatial geometry. Since these are known 
to contribute equally to the deflection, it can therefore 
be stated that the equation of rays will give a deflection 
of only half of the correct value. 

This result is to be expected on optical grounds, because 
the non-Euclidean character of the spatial geometry has 
been neglected in optics. In order to compensate for the 
change in light path, one may use the notion of optical path. 
The optical path represents the distance light travels in a 
vacuum in the same time it travels a distance in the medium. 
If a light ray travels in a medium with spherical symmetry, 
the optical path is given by integral over $n(r)dr$. This means 
that the radial interval of integration must be corrected by 
multiplication with $n(r)$ to take into account the difference 
in path length. Upon integration over $n(r)dr$ instead of the 
original integration over $dr$, it would yield a result in 
which the effect arising from the difference in path length is 
taken into consideration. Using the optical path to correct 
the change in light path, the equation of rays is modified to 
\begin{equation} 
\triangle\theta=\int\biggl[\frac{n^2(r)r^2}{n^2(r_0)r_0^2}-1 
\biggr]^{-1/2}\frac{n(r)dr}{r}. 
\end{equation}
The equation of rays so modified is in complete 
agreement in form with (10). From the proposition which 
has just been proved, one may picture what it is to be 
a curved space in a region of strong gravitational potential. 
When viewed from the present point, the curvature of 
the physical space in the gravitational field of the Sun 
can best be understood in terms of the medium with 
spherical symmetry in which the path of rays is to 
be curved. 

A comparison of (10) with (14) identifies $g_{rr}(r)$ 
with $n^2(r)$. In the geometrical-optics equation of 
rays $n^2(r)$ plays exactly the same role $g_{rr}(r)$ 
has played in the geodesic equation of general relativity. 
This suggests introducing an optical metric tensor $n^2(r)$ in place of $g_{rr}$. The optical metric is an optical version of the Schwarzschild metric, satisfying the eikonal equation with the line element
\begin{equation} 
ds^2=\frac{c^2}{n^2(r)}dt^2=n^2(r)dr^2+r^2d\theta^2. 
\end{equation}
In terms of the components the line element is written as
\begin{equation} 
ds^2=c^2\biggl[1-\frac{GM}{c^2r}-\frac{\triangle P}{c^2\rho}\biggr]^2dt^2=\biggl[1-\frac{GM}{c^2r}-\frac{\triangle P}{c^2\rho}\biggr]^{-2}dr^2+r^2d\theta^2. 
\end{equation}   
It should be noted that the optical metric is based on the interpretation of the deflection of light as a refraction in the gaseous layers surrounding the Sun. 

\section{Radar Echo Delay} 

Shapiro (1964) suggested a fourth test of general relativity. This test involves measuring the time delays between transmission of radar signals from Earth to either Mercury or Venus and detection of the echoes (Shapiro et al. 1968, 1971). The time delay is a result of the path of a radar wave deflected with its varying velocity by the Sun. In content, it amounts to the test of measuring the deflection of light in terms of the time of propagation. According to Weinberg, the time required for light to go from $r_0$ to $r'$ is given by 
\begin{equation} 
\triangle t=\int\limits_{r_0}^{r'} 
\biggl[1-\frac{g_{00}(r)r_0^2}{g_{00}(r_0)r^2}\biggr]^{-1/2} 
\biggl(\frac{g_{rr}(r)}{g_{00}(r)}\biggr)^{1/2}\frac{dr}{c}. 
\end{equation} 
The integral can be evaluated by expanding in the small 
parameters $GM/c^2r$ and $GM/c^2r_0$ to first order, giving 
240 $\mu$sec for the maximum round-trip excess time delay. 

We now have the task to derive an optical form of 
expression for the excess time delay. Its explicit form 
will follow from an equation which specifies the path of 
rays. We can thus obtain the desired expression by converting 
the equation of rays into an equation for the path of rays. 

A procedure starts from (11). Substitution of (11) into (12) 
gives 
\begin{equation} 
\frac{n(r)r^2d\theta}{(dr^2+r^2d\theta^2)^{1/2}}=\mbox{constant}. 
\end{equation} 
Since the path of rays is $ds=(dr^2+r^2d\theta^2)^{1/2}$ in the 
polar coordinates of plane curve, this may also be written as 
\begin{equation} 
\frac{n(r)r^2d\theta}{ds}=\mbox{constant}. 
\end{equation} 
Solving for $ds$, we have 
\begin{equation} 
ds=\frac{n(r)r^2d\theta}{n(r_0)r_0}. 
\end{equation} 
Bouguer's formula has been used in the above equation. By 
making use of the integral in (13), the variable of integration can 
be changed from $d\theta$ to $dr$, obtaining the result 
\begin{equation} 
\triangle s=\int\biggl[1-\frac{n^2(r_0)r_0^2}{n^2(r)r^2} 
\biggr]^{-1/2}dr. 
\end{equation} 
Hence, by dividing $ds$ by $c'(r)=c/n(r)$, the time of 
propagation of rays is found to be 
\begin{equation} 
\triangle t=\int\biggl[1-\frac{n^2(r_0)r_0^2}{n^2(r)r^2} 
\biggr]^{-1/2}\frac{n(r)dr}{c}, 
\end{equation} 
where $c'(r)$ is the speed of propagation of light in a medium 
with spherical symmetry. Although the details are altered by 
the new form of expression, the optical characteristics of 
(22) remain the same as in (13). For the correct calculation 
of excess time delay, therefore, we must take into account a 
difference in the radial interval of the path of rays. 

As discussed in the case of deflection, this requires integrating 
the resulting equation along the optical path. However, it draws 
a clear distinction between geometrical optics and general 
relativity. This is because (22) has already manifested the form 
of an integral over optical path. The geodesic equation of 
general relativity enables us to make a correction not only in 
the velocity of light but also in its path length. If we make 
a correction to the radial component of the path of rays, 
the integral in (22) becomes 
\begin{equation} 
\triangle t=\int\biggl[1-\frac{n^2(r_0)r_0^2}{n^2(r)r^2} 
\biggr]^{-1/2}\frac{n^2(r)dr}{c}. 
\end{equation} 
This integral is in complete agreement in structure with 
the geodesic equation expressed in (17). It becomes evident 
that the equation for rays in geometrical optics also predicts 
the radar echo delay in exactly the same form as given by the 
geodesic equation of general relativity. Again, we identify 
$n^2(r)$ with $g_{rr}(r)$ in their roles, leading to 
consider $n^2(r)$ as an optical version of $g_{rr}(r)$.

\section{Plasma Effect of Corona} 

In radio astronomy, it is possible to measure the deflection 
of radio signals by the Sun with potentially far greater 
accuracy than is possible in optical astronomy. At radio 
frequencies, however, it is necessary to analyze the data 
in terms of a model, in which part of deflection arises from 
general relativity, and the rest is produced by the corona. 
No prediction can be drawn from general relativity since the 
plasma effect is frequency dependent and the gravitational 
effect is not. In contrast with this, the optical approach 
affords a straightforward way to calculate in the integrated 
form the gravitational effect and the plasma effect of corona. 
We are going to evaluate the plasma effect of corona. 

In the discussion of gravitational redshift we have used the notion of a spherically varying refractive index. If we include a frequency-dependent dielectric constant $\epsilon(\omega)$, we can describe in an integrated form the frequency dependence of radio waves. In high-frequency limit, the dielectric constant takes on a simple form. Since $n(r)$ is a varying refractive index in a given medium due to the gradient force, the integrated index of refraction would therefore be written
\begin{equation} 
n^2(r,\omega)=\epsilon(\omega)n^2(r)= 
\biggl(1-\frac{4\pi e^2N}{m\omega^2}\biggr) 
\biggl[1+\frac{1}{c^2}\biggl(\triangle\phi-\frac{\triangle P} 
{\rho}\biggr)\biggr]^{-2}, 
\end{equation} 
where $m$ and $e$ are the mass and charge on the electron, 
and $N$ is the total number of electrons per unit volume. Since 
the characteristics of propagation obviously depend on the 
index of refraction, it is very natural to expect 
the frequency dependence of the deflection of light so discussed. 
In fact, there has been an interferometric measurement of the deflection of radio waves using an effective index of refraction. Muhleman et al. (1970) analyzed their experimental 
data by using geometrical-optics techniques in a spherically 
symmetric refracting medium of index
\begin{equation} 
n(r,\omega)=1+\frac{2GM}{c^2r}-\frac{2\pi e^2N(r)}{m\omega^2},
\end{equation} 
where $N(r)$ is the electron-density profile in the corona 
and interplanetary medium. We may draw a correspondence between their technique and the present approach. 

As a first important example, we consider the deflection of 
light as a combination of the general relativistic effect 
and of refraction in the coronal plasma. The expected angular 
deviation can be accurately computed using the 
frequency-dependent refractive index expressed in (24) in 
the equation of rays (14): 
\begin{equation} 
\triangle\theta=\int\biggl[\frac{n^2(r,\omega)r^2} 
{n^2(r_0,\omega)r_0^2}-1\biggr]^{-1/2}\frac{n(r,\omega)dr}{r}. 
\end{equation} 
In order to evaluate this integral, we use in the integrand 
expansions in the small parameters. It is both easier and more 
instructive to evaluate the integral after the expansions. The 
integration of (26) can be carried to first order in the small 
parameters with high accuracy. 

The argument of the square root in (26) can be expanded to 
first order in the small parameters. Taking only the leading 
gravitational potential in (7), we have 
\begin{eqnarray} 
\frac{n^2(r,\omega)r^2}{n^2(r_0,\omega)r_0^2}-1 &\simeq& 
\frac{r^2}{r_0^2}\biggl[1+\frac{2GM}{c^2}\biggl(\frac 1r - 
\frac{1}{r_0}\biggr)-\frac{4\pi e^2}{m\omega^2}\bigl(N(r)- 
N(r_0)\bigr)\biggr]-1 \nonumber \\&\simeq& 
\biggl(\frac{r^2}{r_0^2}-1\biggr)\biggl[1-\frac{2GMr}{c^2r_0(r+r_0)}+ 
\frac{4\pi e^2r^2}{m\omega^2(r_0^2-r^2)}\bigl(N(r)-N(r_0)\bigr)\biggr], 
\end{eqnarray} 
so (26) gives 
\begin{equation} 
\triangle\theta\simeq\int\biggl(\frac{r^2}{r_0^2}-1\biggr)^{-1/2} 
\frac{dr}{r}\biggl[1+\frac{GM}{c^2r}+\frac{GMr}{c^2r_0(r+r_0)}- 
\frac{2\pi e^2N(r)}{m\omega^2}- 
\frac{2\pi e^2r^2\bigl(N(r)-N(r_0)\bigr)}{m\omega^2(r_0^2-r^2)}\biggr]. 
\end{equation} 
Consequently, the deflections from the individual effects 
are combined linearly. Refraction effect in the solar corona 
is now represented by 
\begin{equation} 
\delta\theta_c\simeq\int\biggl(\frac{r^2}{r_0^2}-1\biggr)^{-1/2} 
\frac{dr}{r}\biggl[\frac{2\pi e^2N(r)}{m\omega^2}+ 
\frac{2\pi e^2r^2\bigl(N(r)-N(r_0)\bigr)}{m\omega^2(r_0^2-r^2)}\biggr]. 
\end{equation} 
This must be an addition to the general-relativity deflection. 

In the Allen-Baumbach model (Allen 1947), the electron-density profile in the corona is assumed to have the form $N(r)=1.55\times 10^8 (R_{\odot}/r)^6$ electron/cm$^3$. Using more recent results on the corona, Erickson (1964) found that $N(r)=5\times 10^5(R_{\odot}/r)^2$ electron/cm$^3$ represents the data reasonably well from $4R_{\odot}$ to $20R_{\odot}$. Refraction effect is significant where $r<3R_{\odot}$, at which the 
$(R_{\odot}/r)^6$ term dominates. Hence, we use the 
electron distribution of the Allen-Baumbach model, resulting in 
\begin{equation} 
\delta\theta_c\simeq\frac{6.24\times10^{15}}{f^2} 
\int\biggl(\frac{r^2}{r_0^2}-1\biggr)^{-1/2}\frac{dr}{r}\biggl[ 
\biggl(\frac{R_{\odot}}{r}\biggr)^6+\frac{r^2}{r_0^2-r^2}\biggl( 
\biggl(\frac{R_{\odot}}{r}\biggr)^6-\biggl(\frac{R_{\odot}}{r_0} 
\biggr)^6\biggr)\biggr]. 
\end{equation} 
The integration for $\delta\theta_c$ is straightforward, 
and gives 
\begin{equation} 
\delta\theta_c\simeq\frac{6.24\times10^{15}}{f^2} 
\biggl(\frac{R_{\odot}}{r_0}\biggr)^6\biggl[\frac{105}{48}\theta+ 
\frac{57}{48}\cos\theta\sin\theta+\frac{11}{24} 
\cos^3\theta\sin\theta+\frac 16\cos^5\theta\sin\theta\biggr], 
\end{equation} 
where $\cos\theta=r_0/r$. 

The total change in $\theta$ as $r$ decreases from infinity 
to its minimum value $r_0$ and then increases again to infinity 
is just twice its change from $\infty$ to $r_0$, that is, 
$2\triangle\theta$. Hence, the deflection of the path of rays 
from a straight line is given by 
$\delta\theta=2\triangle\theta-\pi$, which is calculated 
positively if concave toward the Sun and negatively if convex. 
Putting in the numerical factors, the total deflection is 
\begin{equation} 
\delta\theta\simeq 1.75''\biggl(\frac{R_{\odot}}{r_0}\biggr)- 
\frac{6.24\times 10^{15}}{f^2}\biggl(\frac{105\pi}{48}\biggr) 
\biggl(\frac{R_{\odot}}{r_0}\biggr)^6. 
\end{equation} 

Equation (32) describes an interesting behaviour of the radiation bending near the Sun. The first term represents the general relativistic effect by which the path of rays is bent toward the Sun. The second term represents the coronal refraction by which the path of rays is bent away from the Sun to the contrary. This explanation is given by the difference in sign between those terms. At optical frequencies, coronal refraction is extremely small, so it can be neglected. However, it plays an important part at radio frequencies, as can be seen when we illustrate the deflection angle as a function of frequency for the distances in solar radii of the ray's point of closest approach to the Sun's center. 

The question might be raised as to whether varying velocity 
of light in the coronal plasma also gives rise to a change 
in path length therein. If we assume that varying velocity 
of light in the coronal plasma does not give rise to a change 
in the path length of rays, the radial interval of 
integration must still be corrected by multiplication with 
$n(r)$ even in the coronal plasma, not with $n(r,\omega)$ 
as used in (26). We must then drop the fourth term in the 
integrand of integral in (28), that is, the first term in (29). 
Coronal refraction thus obtained will be exactly the same 
as what one finds by evaluating the original equation of 
rays (13) on purely optical grounds. Note that under such assumption there is a complete agreement in the form of expression for the plasma effect between (13) and (26). In fact, the evaluation of coronal refraction from the equation of rays (13) was carried out to first order by Bracewell et al. (1969). Their calculation gives $82(R_{\odot}/r_0)^6$ sec for the angular deviation of a ray of frequency 9.6 GHz in the corona assuming the Allen-Baumbach model. When Erickson's coronal model is instead assumed, the angular deviation is 
given by $0.14(R_{\odot}/r_0)^2$ sec. Seielstad et al. (1970) used in data analysis these values as parameters 
describing refraction effects in the solar corona, when 
they measured the deflection of 9.602 GHz radiation from 
3C279 in the solar gravitational field using an interferometer 
at the Owens Valley Radio Observatory. The results of 
their calculation are in exact agreement with what we 
would obtain from each model if we excluded the first term 
from the integrand of integral in (29). However, on the assumption that varying velocity of light in the coronal plasma also gives rise to a change in its path length, we obtain from (32) coronal refraction of $96(R_{\odot}/r_0)^6$ sec for 9.6 GHz frequency. If we used Erickson's coronal model, we would obtain coronal refraction of $0.21(R_{\odot}/r_0)^2$ sec. 

As characteristic for a comparison, one may write down the 
difference of calculation in terms of the metric tensor. 
We have used in (32) the metric tensor of the components 
\begin{equation} 
g_{00}=\frac{1}{\epsilon(\omega)n^2(r)}\quad\mbox{and}\quad 
g_{rr}=\epsilon(\omega)n^2(r). 
\end{equation} 
When viewed from the present point, their calculation 
corresponds to the case which is obtained when the components 
of the metric tensor are 
\begin{equation} 
g_{00}=\frac{1}{\epsilon(\omega)n^2(r)}\quad\mbox{but}\quad 
g_{rr}=1\mbox{ or }n^2(r). 
\end{equation} 
The reason for this difference is readily understood by 
referring to the equations of rays (14) and (13) from which 
angular deviations were calculated respectively. 

As a second example, let us calculate the plasma effect of 
corona on the time delay of radar echoes. The time of propagation of rays is given by (23). To evaluate its frequency dependence, 
the integration of (23) should be carried out with the 
frequency-dependent refractive index in (24): 
\begin{equation} 
t(r_0,r')=\int\limits_{r_0}^{r'}\biggl[1- 
\frac{n^2(r_0,\omega)r_0^2}{n^2(r,\omega)r^2}\biggr]^{-1/2} 
\frac{n^2(r,\omega)dr}{c}. 
\end{equation} 

In order to evaluate this integral, we once again use in the 
integrand the expansions in the small parameters to first order. 
Proceeding in exactly the same way as for (26), (35) gives 
\begin{equation} 
t(r_0,r')\simeq\int\limits_{r_0}^{r'}\biggl(1-\frac{r_0^2}{r^2} 
\biggr)^{-1/2}\frac{dr}{c}\biggl[1+\frac{2GM}{c^2r}+ 
\frac{GMr_0}{c^2r(r+r_0)}-\frac{4\pi e^2N(r)}{m\omega^2}- 
\frac{2\pi e^2r_0^2\bigl(N(r_0)-N(r)\bigr)} 
{m\omega^2(r^2-r_0^2)}\biggr]. 
\end{equation} 
The time required for radar signals to travel to Mercury and 
be reflected back to Earth is $2[t(r_E,r_0)+t(r_0,r_M)]$, 
where $r_E$ and $r_M$ are astronomical radii of the orbits 
of the Earth and the Mercury around the Sun. The round-trip 
excess time delay is then given by 
$\delta t=2[t(r_E,r_0)+t(r_0,r_M)-T(r_E,r_0)-T(r_0,r_M)]$, 
where $T(r_E,r_0)$ and $T(r_0,r_M)$ are the times required 
for radar signals to travel the paths in straight lines at speed 
$c$. The distance $r_0$ of closest approach of the radar wave 
to the center of the Sun is much smaller than the distances 
$r_E$ and $r_M$ of the Earth and Mercury from the Sun. 

\begin{figure}
\includegraphics[scale=0.5]{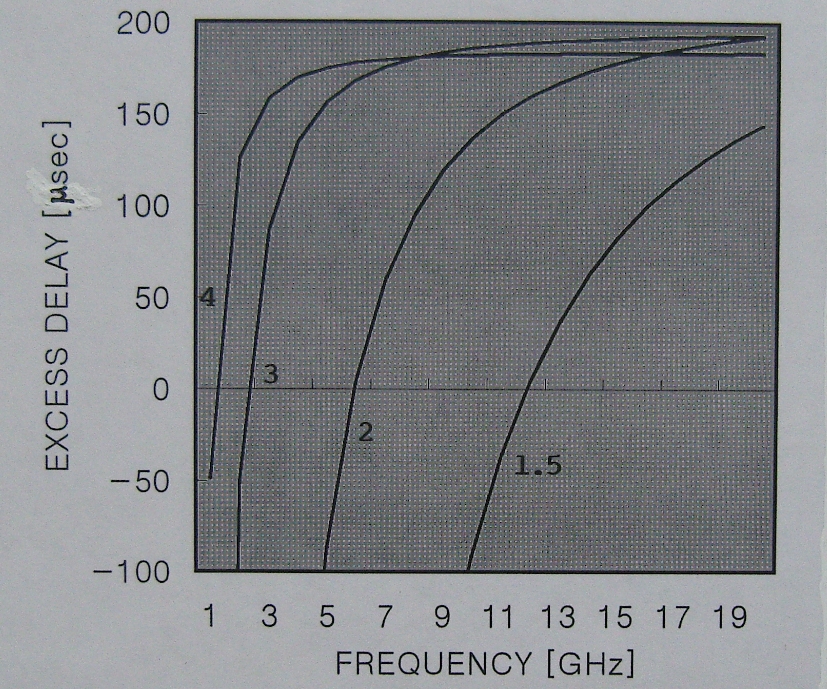} 
\caption{excess time delay as a function of frequency for $r_0/R_{\odot}= 1.5, 2, 3, 4$}
\end{figure}

Assuming the electron distribution of the Allen-Baumbach 
model, the integral yields 
\begin{equation} 
\delta t\simeq\frac{4GM}{c^3}\biggl[1+\ln\biggl( 
\frac{4r_Er_M}{r_0^2}\biggr)\biggr]- 
\frac{6.24\times 10^{15}}{f^2}\biggl(\frac{21\pi r_0}{4c}\biggr) 
\biggl(\frac{R_{\odot}}{r_0}\biggr)^6. 
\end{equation} 
If, instead, we use Erickson's coronal model, we then have 
\begin{equation} 
\delta t\simeq\frac{4GM}{c^3}\biggl[1+\ln\biggl( 
\frac{4r_Er_M}{r_0^2}\biggr)\biggr]- 
\frac{2.01\times 10^{13}}{f^2}\biggl(\frac{6\pi r_0}{c}\biggr) 
\biggl(\frac{R_{\odot}}{r_0}\biggr)^2. 
\end{equation}
As in the case of deflection, because of the sign difference, the plasma effect of the solar corona on the time delay is opposite to what is usually expected from the general relativistic effect. For either equation, the positive terms on the right describe a general relativistic delay in the time it takes a radar signal to travel to Mercury and back. In contrast, the negative terms describe a coronal plasma contraction, that is, a radar time contraction. Figure 1 shows calculated curves of the excess time delay as a function of frequency for $r_0/R_{\odot}=1.5,2,3,4$. Exactly the same curves are obtained for the deflection angle calculated from (32). In the case of deflection angle, the ordinate ranges from $-5$ second to $+1$ second.

We now compare Shapiro's calculation with the results 
obtained from the preceding equations. At the beginning 
of measurement, Shapiro estimated $\delta t\simeq 
140-370\,\mu$sec for observations of Mercury near 
superior conjunction with $r_0 \approx 4R_{\odot}$ 
at 430 MHz frequency of the Arecibo Ionospheric Observatory. 
This difference in time delays between the general relativistic 
effect and the coronal plasma effect was nowhere large 
enough and positive for a really reliable result to be obtained 
solely from Arecibo data. He was thus tried to reduce the 
plasma effect by a factor of almost 400 by using measurements 
made at 8350 MHz frequency of Haystack radar of Lincoln 
Laboratory. For observation of Mercury with $r_0=4R_{\odot}$ 
at 430 MHz radiation, (37) gives 
$\delta t\simeq 180-1260\,\mu$sec, and (38) gives 
$\delta t\simeq 180-1190\,\mu$sec. The values obtained 
for the plasma effect are about three times larger than Shapiro's 
estimate. This is because we have made a correction in the 
radial interval of the path and integrated the resulting 
equation along the optical path bending near the Sun, 
unlike Shapiro's calculation along the optical 
path in a straight line without any correction. If we 
assume that the difference in path length is ascribed solely 
to differing gravitational potential, the radial interval 
of integration must still be corrected by multiplication 
with $n(r)$ even in the coronal plasma, not with $n(r,\omega)$ 
as used in (35). The excess time delays are then given by 
$180-840\,\mu$sec and $180-790\,\mu$sec, respectively. 
The values so obtained for the plasma effect are just what 
we should expect if the excess time delays were calculated 
from the equation of rays without any correction on purely 
optical grounds.

I conclude this section by commenting on the current evaluation 
of the plasma effect in the corona. A refraction effect in the 
corona has previously been estimated using the equation 
of rays in literatures. But the equation of rays is lacking a correction in the radial interval of the path of rays. So the current estimate corresponds physically to half the correct value. The refraction effect in the corona should be evaluated by the modified equation of rays (14). In the case of excess time delay, the correct calculation can now be made using (23). It is desirable and required to use the correct value as a parameter in radio data analysis for a least-squares fit.

\section{Concluding Remarks}
 
When viewed from the optical point, the redshift is no more than an effect that the velocity of light is altered, linearly to a first approximation, by the medium in the gravitational field. This leads us to consider the redshift effect as due ultimately to the pressure gradient force. It is because gravitational potential is a leading term of the pressure gradient force in the medium of propagation. Not only gravitational potential, accordingly, but any change in pressure and density of the medium would give rise to an effect of the same kind on the redshift of spectral lines. In fact, asymmetry profiles observed in the redshifts of solar spectral lines seem to be of pressure character at the point of observation on the solar disk. Einstein's interpretation is in a certain sense the general-relativity analogue of the Lorentz time dilation in special relativity physics. I should mention the Lorentz time dilation. In my argument (Yi 1997, 2000), which shows a physics behind the aberration of starlight, the Lorentz time dilation is a result of confusion leading us to a misunderstanding of relativistic phenomena.

There is an agreement in structure between the equation 
for rays in geometrical optics and the geodesic equation of 
general relativity. From a comparison with the geodesic equation, 
we learn that the radial component of the path of rays must take the form of an optical path in the equation of rays. To be 
reconciled with general relativity, that is, the length of the path of a ray should be given by $ds^2=n^2(r)dr^2+r^2d\theta^2$, 
not by $ds^2=dr^2+r^2d\theta^2$, in a medium with spherical 
symmetry of refractive index $n(r)$. Using this path length in 
the equation of rays, we find a complete agreement in structure 
between the equation of rays and the geodesic equation. 
Comparisons have been made with the geodesic equations 
formulated by Weinberg. In light of this fact, Weinberg's 
formalism has opened a door to introduce the relation 
between general relativity and geometrical optics, apparently 
unrelated areas of physics. Indeed, it was an agreement in 
form between (10) and (13) that enabled the present approach 
to be proposed.

The development of an optical analogy suggests introducing 
$n^2(r)$ in place of $g_{rr}$ as an optical version of the 
Schwarzschild metric. While $g_{rr}$ is a solution for the motion of a particle in the gravitational field, $n(r)$ is a refractive index given by properties of the medium of propagation. The equation of rays provides a theoretical curve for observation that any beam of radiation is deflected during its passage near the Sun as a result of the gravitational effect and of refraction in the coronal electron plasma. In addition to the gravitational effect, there would be a pressure effect in the gaseous layers surrounding the Sun. As the light rays are passed close to the Sun's disk, a pressure effect would appear pronounced in addition to the gravitational effect. In the case of a conducting medium, we may expect even a magnetic-field effect from the hydrodynamic equation. The optical approach may be supposed to serve as a means of describing additional gravitational effects involving optical phenomena. But it should be emphasized that the optical point of view replaces the general-relativity explanations in terms of time and gravitation.

\bigskip

\noindent{\bf Acknowledgements} I should like to express my gratitude to a reviewer of this Journal for comments and corrections. Although a long time has passed, I wish to express gratitude to a reviewer of Physics Essays for corrections in the early draft of this paper.

\bigskip

\noindent Allen, C. W.: Mon. Not. R. Astron. Soc. 
{\bf 107}, 426 (1947).
\vspace{2pt}

\noindent Blamont, J. E., Roddier, F.: Phys. Rev. Lett. 
{\bf 7}, 437 (1961).
\vspace{2pt}

\noindent Born, M., Wolf, E.: Principles of Optics. 5th ed. p. 123. Pergamon Press, Oxford (1975).
\vspace{2pt}

\noindent Bracewell, R. N., Eshelman, V. R., Hollweg, J. V.:
Astrophys. J. {\bf 155}, 367 (1969). 
\vspace{2pt}

\noindent De Felice, F.: Gen. Rel. Grav. {\bf 2}, 347 (1971).
\vspace{2pt}

\noindent Eddington, A. S.: Space, Time and Gravitation. p. 109. Cambridge University Press, Cambridge (1920).
\vspace{2pt}

\noindent Einstein, A.: Ann. Physik {\bf 35}, 898 (1911).
\vspace{2pt}

\noindent Erickson, W. C.: Astrophys. J. {\bf 139}, 
1290 (1964).
\vspace{2pt}

\noindent Fabry, Ch., Boisson, H.: Comptes Rendus {\bf 148},
688 (1909).
\vspace{2pt}

\noindent Hay, H. J., Schiffer, J. P., Cranshaw, T. E., 
Egelstaff, P. A.: Phys. Rev. Lett. {\bf 4}, 165 (1960).
\vspace{2pt}

\noindent Higgs, L. A.: Mon. Not. R. Astron. Soc. 
{\bf 121}, 421 (1960).    
\vspace{2pt}

\noindent Jewell, L. F.: J. de Phys. {\bf 6}, 84 (1897). 
\vspace{2pt}

\noindent Muhleman, D. O., Ekers, R. D., Fomalont, E. B.:
Phys. Rev. Lett. {\bf 24}, 1377 (1970). 
\vspace{2pt}

\noindent Plebanski, J.: Phys. Rev. {\bf 54}, 1396 (1960).
\vspace{2pt}

\noindent Pound, R. V., Rebka Jr., G. A.: Phys. Rev. Lett. 
{\bf 3}, 439 (1959).
\vspace{2pt}

\noindent Pound, R. V., Rebka Jr., G. A.: Phys. Rev. Lett.
{\bf 4}, 337 (1960).
\vspace{2pt}

\noindent Schiff, L. I.: Am. J. Phys. {\bf 28}, 340 (1960).
\vspace{2pt}

\noindent Seielstad, G. A., Sramek, R. A., Weiler, K. W.:  
Phys. Rev. Lett. {\bf 24}, 1373 (1970).
\vspace{2pt}

\noindent Shapiro, I. I.: Phys. Rev. Lett. {\bf 13}, 789 
(1964).
\vspace{2pt}

\noindent Shapiro, I. I., et al.: Phys. Rev. Lett. {\bf 20}, 
1265 (1968).
\vspace{2pt}

\noindent Shapiro, I. I., et al.: Phys. Rev. Lett. 
{\bf 26}, 1132 (1971).
\vspace{2pt}

\noindent Weinberg, S.: Gravitation and Cosmology. p. 189. John Wiley \& Sons, New York (1972). 
\vspace{2pt}

\noindent White, H. E.: Introduction to Atomic Spectra. p. 431. McGraw-Hill, Tokyo (1934).
\vspace{2pt}

\noindent Yi, Y. G.: Physics Essays {\bf 10}, 186 (1997).
\vspace{2pt}

\noindent Yi, Y. G.: arXiv:physics/0006005 (2000). 
\vspace{2pt}

\end{document}